\documentclass[pra,aps,twocolumn]{revtex4}
\usepackage{epsf}
\usepackage{amsmath}
\usepackage{amssymb}

\begin{document}

\title{A comparative study of relative entropy of entanglement,
concurrence and negativity\footnote{published in {\em J. Opt. B:
Quantum Semiclass. Opt.} {\bf 6} (2004) 542 –- 548; online at {\tt
stacks.iop.org/JOptB/6/542}}}

\author{Adam Miranowicz and Andrzej Grudka}
\date{\today}

\affiliation{Faculty of Physics, Adam Mickiewicz University, 61-614
Pozna\'n, Poland}

\begin{abstract}
The problem of ordering of two-qubit states imposed by relative
entropy of entanglement (E) in comparison to concurrence (C) and
negativity (N) is studied. Analytical examples of states
consistently and inconsistently ordered by the entanglement
measures are given. In particular, the states for which any of the
three measures imposes order opposite to that given by the other
two measures are described. Moreover, examples are given of pairs
of the states, for which (i) N'=N'' and C'=C'' but E' is different
from E'', (ii) N'=N'' and E'=E'' but C' differs from C'', (iii)
E'=E'', N'$<$N'' and C'$>$C'', or (iv) states having the same E, C,
and N but still violating the Bell-Clauser-Horne-Shimony-Holt
inequality to different degrees.
\\


\end{abstract}


\maketitle

\pagenumbering{arabic}

\section{Introduction}

Quantum entanglement is a key resource for quantum information
processing but still its mathematical description is far from
completeness \cite{horodecki-book} and its properties are more and
more intriguing. In particular, Eisert and Plenio \cite{eisert}
five years ago observed by Monte Carlo simulation of pairs of
two-qubit states $\sigma'$ and $\sigma''$ that entanglement
measures (say $E^{(1)}$ and $E^{(2)}$) do not necessarily imply the
same ordering of states. This means that the intuitive requirement
\begin{equation}
E^{(1)}(\sigma') < E^{(1)}(\sigma'') \Leftrightarrow
E^{(2)}(\sigma') < E^{(2)}(\sigma'') \label{N01}
\end{equation}
can be violated. The problem was then analyzed by others
\cite{zyczkowski99,virmani,verstraete,zyczkowski02,wei1,wei2,miran1,miran2,mg1}.
In particular, Virmani and Plenio \cite{virmani} proved that all
good asymptotic entanglement measures  are either identical or fail
to impose consistent orderings on the set of all quantum states.
Here, an entanglement measure is referred to as `good' if it
satisfies (at least most of) the standard criteria
\cite{vedral97a,vedral98,horodecki00} including that for pure
states it should reduce to the canonical form given by the von
Neumann entropy of the reduced density matrix.

We will study analytically the problem of ordering of two-qubit
states imposed by the following three standard entanglement
measures.

The first measure to be analyzed here is the relative entropy of
entanglement (REE) of a given state $\sigma$, which is defined by
Vedral {\em et al} \cite{vedral97a,vedral98} (for a review see
\cite{vedral02}) as the minimum of the quantum relative entropy
$S(\sigma ||\rho )={\rm Tr}\,( \sigma \lg \sigma -\sigma\lg \rho )$
taken over the set ${\cal D}$ of all separable states $\rho$,
namely
\begin{equation}
E(\sigma)={\rm min}_{\rho \in {\cal D}} S(\sigma ||\rho )= S(\sigma
||\bar{\rho}), \label{N02}
\end{equation}
where $\bar{\rho}$ denotes a separable state closest to $\sigma$.
We assume, for consistency with the other entanglement measures
that $\lg$ stands for $\log_2$ although in the original Vedral {\em
et al} papers \cite{vedral97a,vedral98} the natural logarithms were
chosen. It is usually difficult to calculate analytically the REE
with exception of states with high symmetry, including those
discussed in sections 3 and 4. Thus, in general, the REE is
calculated numerically using the methods described in, e.g.,
\cite{vedral98,doherty,rehacek}. The REE satisfies both continuity
and convexity [monotonicity under discarding information, $E(\sum_i
p_i\sigma_i)\le \sum_i p_iE(\sigma_i)$] \cite{donald}, but it does
not fulfill additivity [$E(\sigma_1\otimes\sigma_2) =E(\sigma_1)+
E(\sigma_2)$] \cite{vollbrecht}.

The second measure of entanglement for a given two-qubit state
$\sigma$ is the Wootters concurrence $C({\sigma})$ defined as
\cite{wootters}
\begin{equation}
C({\sigma})=\max \{0,\lambda_1-\lambda_2-\lambda_3-\lambda_4\},
\label{N03}
\end{equation}
where the $\lambda _{i}$'s  are the square roots of the eigenvalues
of ${\sigma }({\sigma }^{(y)}\otimes {\sigma }^{(y)})\sigma^*({
\sigma }^{(y)}\otimes {\sigma }^{(y)})$ put in nonincreasing order,
${\sigma }^{(y)}$ is the Pauli spin matrix, and asterisk stands for
complex conjugation. The concurrence $C({\sigma})$ is monotonically
related to the entanglement of formation $E_{\rm form}({\sigma})$
\cite{bennett96} as given by the Wootters formula \cite{wootters}
\begin{equation}
E_{\rm form}(\sigma)=
h\left(\textstyle{\frac{1}{2}}[1+\sqrt{1-C^2(\sigma)}]\right)
\label{N04}
\end{equation}
in terms of the binary entropy $h(x)=-x\lg x-(1-x)\lg(1-x)$. The
concurrence and entanglement of formation satisfy convexity
\cite{wootters,horodecki04}. But, to our knowledge, the question
about additivity of the entanglement of formation is still open
\cite{wootters01,horodecki04}.

The third useful measure of entanglement is the negativity --  a
measure related to the Peres-Horodecki criterion \cite{peres} as
defined by
\begin{equation}
N(\sigma)=2\sum_{j}\max(0,-\mu _{j}),  \label{N05}
\end{equation}
where $\mu _{j}$'s are the eigenvalues of the partial transpose
$\sigma^{\Gamma}$ of the density matrix $\sigma$ of the system.
Note that for any two-qubit states,  $\sigma^{\Gamma}$ has at most
one negative eigenvalue. As shown by Audenaert {\em et al}
\cite{audenaert} and as subsidiarily by Ishizaka \cite{ishizaka04},
the negativity of any two-qubit state $\sigma$ is a measure closely
related to the PPT entanglement cost as follows:
\begin{equation}
E_{\rm PPT}(\sigma)= \lg[N(\sigma)+1], \label{N06}
\end{equation}
which is the cost of the exact preparation of $\sigma$ under
quantum operations preserving the positivity of the partial
transpose (PPT). $E_{\rm PPT}(\sigma)$, similarly to $E_{\rm
form}(\sigma)$ and $E(\sigma)$, gives an upper bound of the
entanglement of distillation \cite{bennett97}. As shown by Vidal
and Werner \cite{vidal}, the negativity is a convex function,
however $E_{\rm PPT}(\sigma)$ is {\em not} convex as a combination
of the convex $N(\sigma)$ and the concave logarithmic function.
Nevertheless, $E_{\rm PPT}(\sigma)$ satisfies additivity. For a
pure state $|\psi_P\rangle$, it holds
$C(|\psi_P\rangle)=N(|\psi_P\rangle)$ but $E_{\rm
PPT}(|\psi_P\rangle)\ge E_{\rm form}(|\psi_P\rangle)$, where
equality holds for separable and maximally entangled states. For
these reasons, we will apply the concurrence and negativity instead
of $E_{\rm form}$ and $E_{\rm PPT}$.

\section{Numerical comparison of state orderings}

In previous works much attention was devoted to the ordering
problem for the concurrence versus the negativity
\cite{eisert,zyczkowski99,miran1,miran2,mg1}. Here, we will study
analytically the ordering of two qubit-states imposed by the REE in
comparison to the other two measures. But first let us show the
violation of condition (\ref{N01}) by numerical simulation. We have
generated `randomly' $10^5$ two-qubit states according to the
method described by \.Zyczkowski {\em et al}
\cite{pozniak,zyczkowski98} and applied, e.g., by Eisert and Plenio
\cite{eisert}. The results are shown in figure 1, where for each
generated state $\sigma$ we have plotted $E(\sigma)$ versus
$C(\sigma)$, $E(\sigma)$ versus $N(\sigma)$, and $N(\sigma)$ versus
$C(\sigma)$. It is worth noting that apparent saw-like irregularity
of distribution of states (along the x-axes) is an artifact
resulting from the modification of the original \.Zyczkowski {\em
et al} method. Namely, we have performed simulations sequentially
in 10 rounds and during the $k$th round we plotted the three
entanglement measures only for those $\sigma$ for which $C(\sigma)$
was greater than $(k-1)/10$. The speed-up of this biased simulation
is a result of fast procedures for calculating the negativity or
concurrence and very inefficient ones for calculating the REE
\cite{vedral98,doherty,rehacek,ishizaka04}. Our sequential method
could be applied since the main goal for generating states was to
check efficiently the boundaries of the depicted regions but not
the distribution of states.
\begin{figure}
\centerline{
 \epsfxsize=4cm\epsfbox{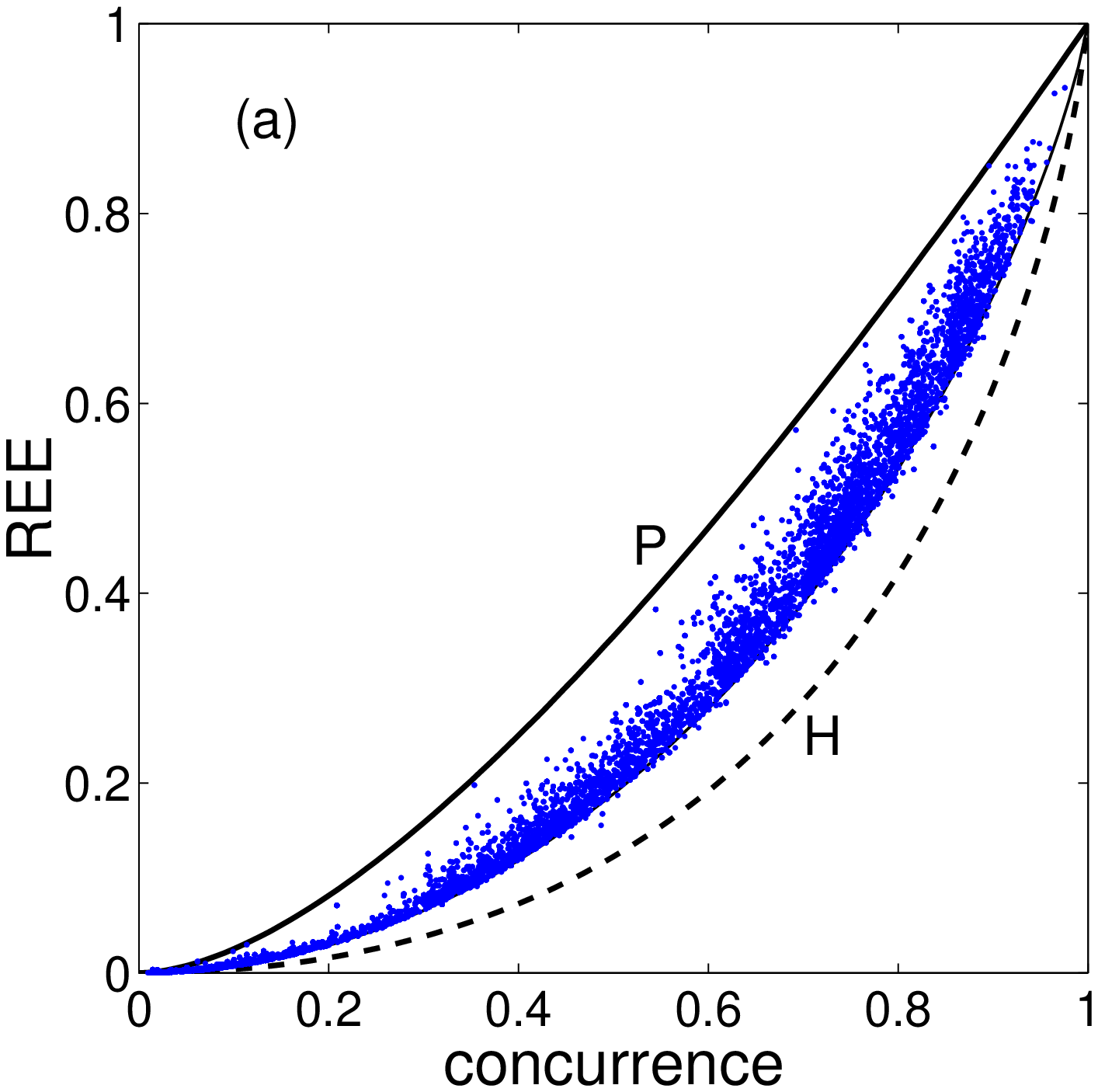}
 \epsfxsize=4cm\epsfbox{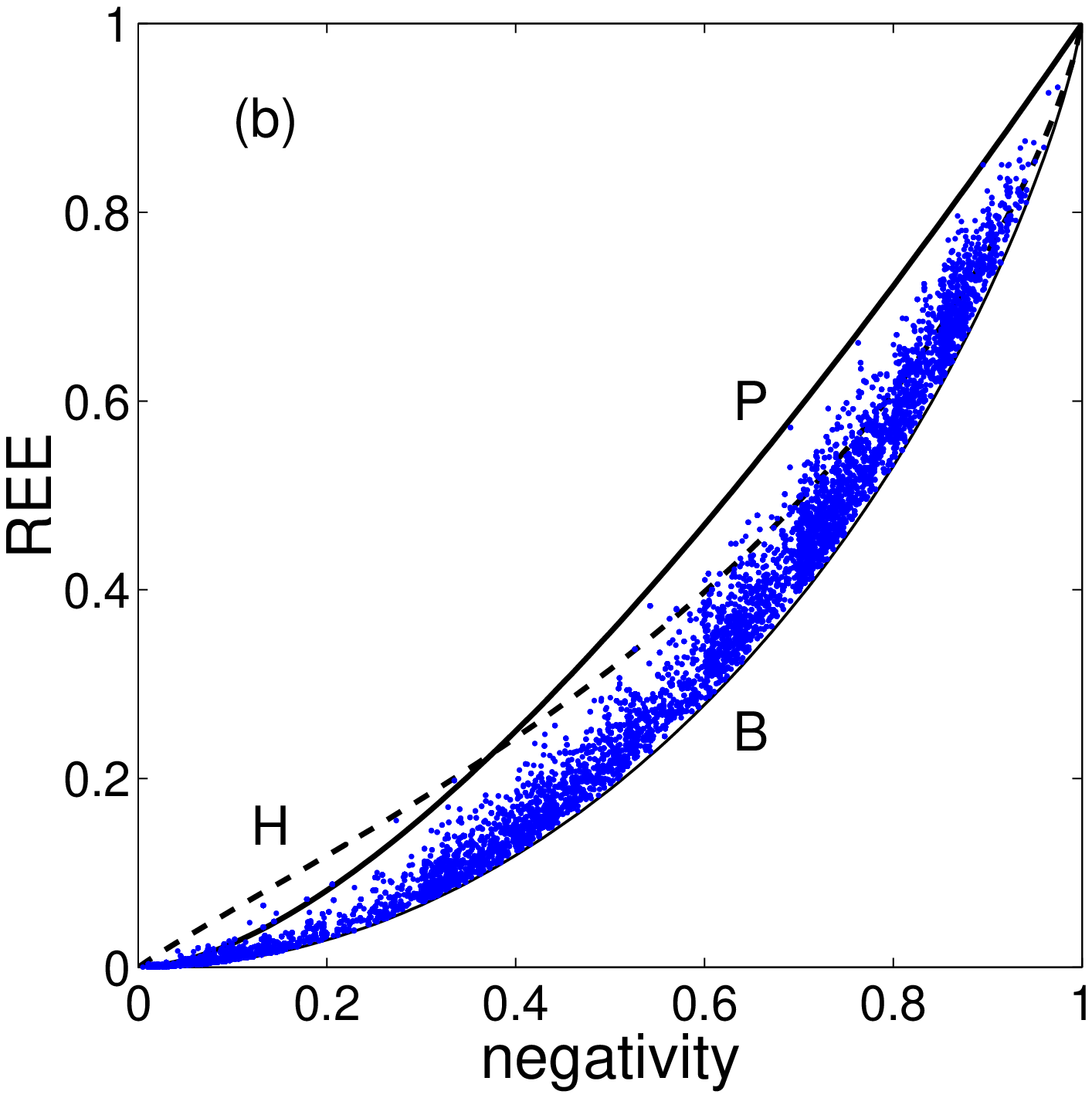}}
 \centerline{\epsfxsize=4cm\epsfbox{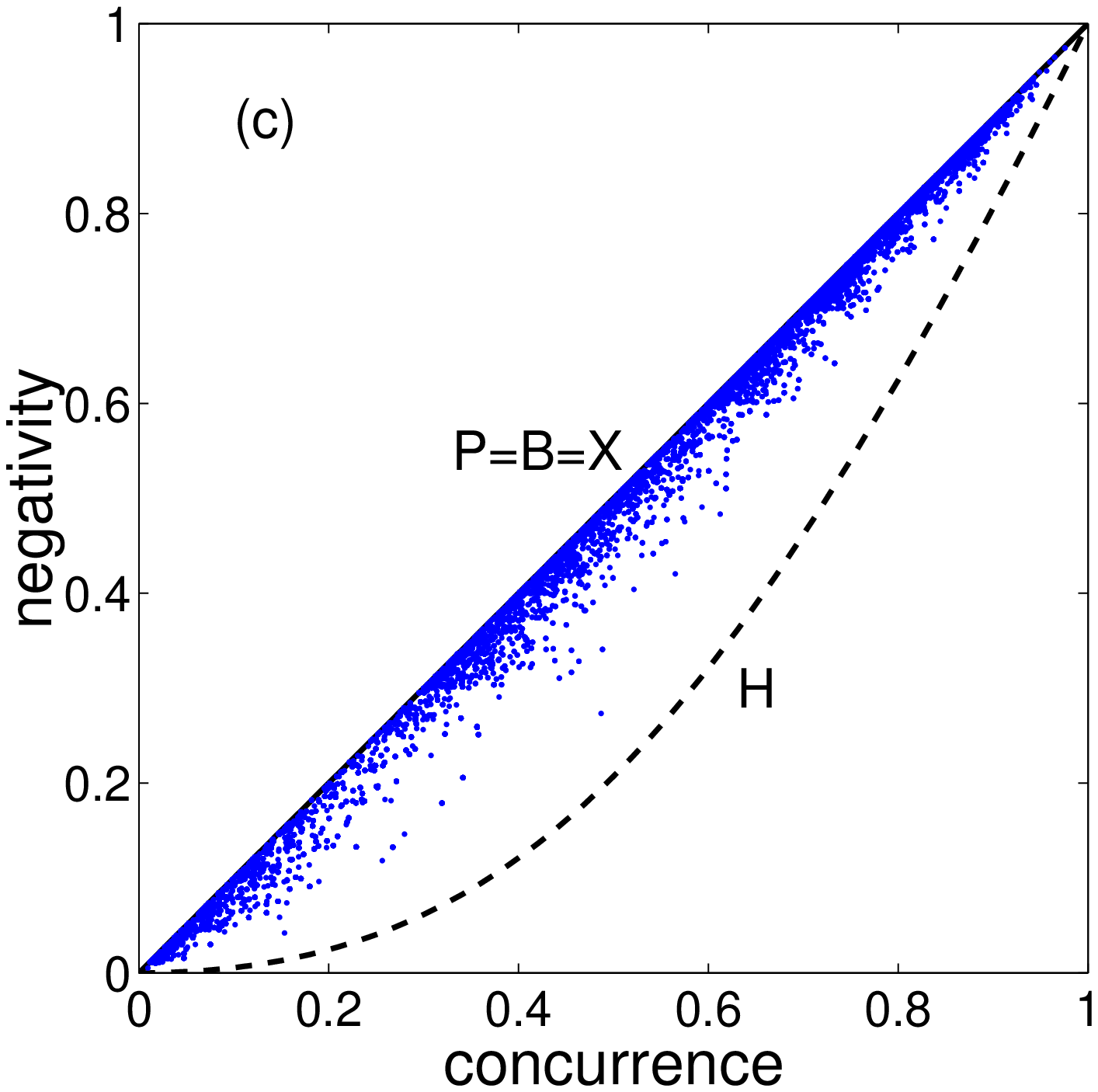}}
\caption{Numerical simulations of about $10^5$ quantum states
$\sigma$: (a) REE $E(\sigma)$ versus concurrence $C(\sigma)$, (b)
$E(\sigma)$ versus negativity $N(\sigma)$, and (c) $N(\sigma)$
versus $C(\sigma)$. Curves correspond to the Horodecki (H), pure
(P), Bell diagonal (B) and $\sigma_X$ (X) states.}
\end{figure}
\begin{figure}
\centerline{
 \epsfxsize=4cm\epsfbox{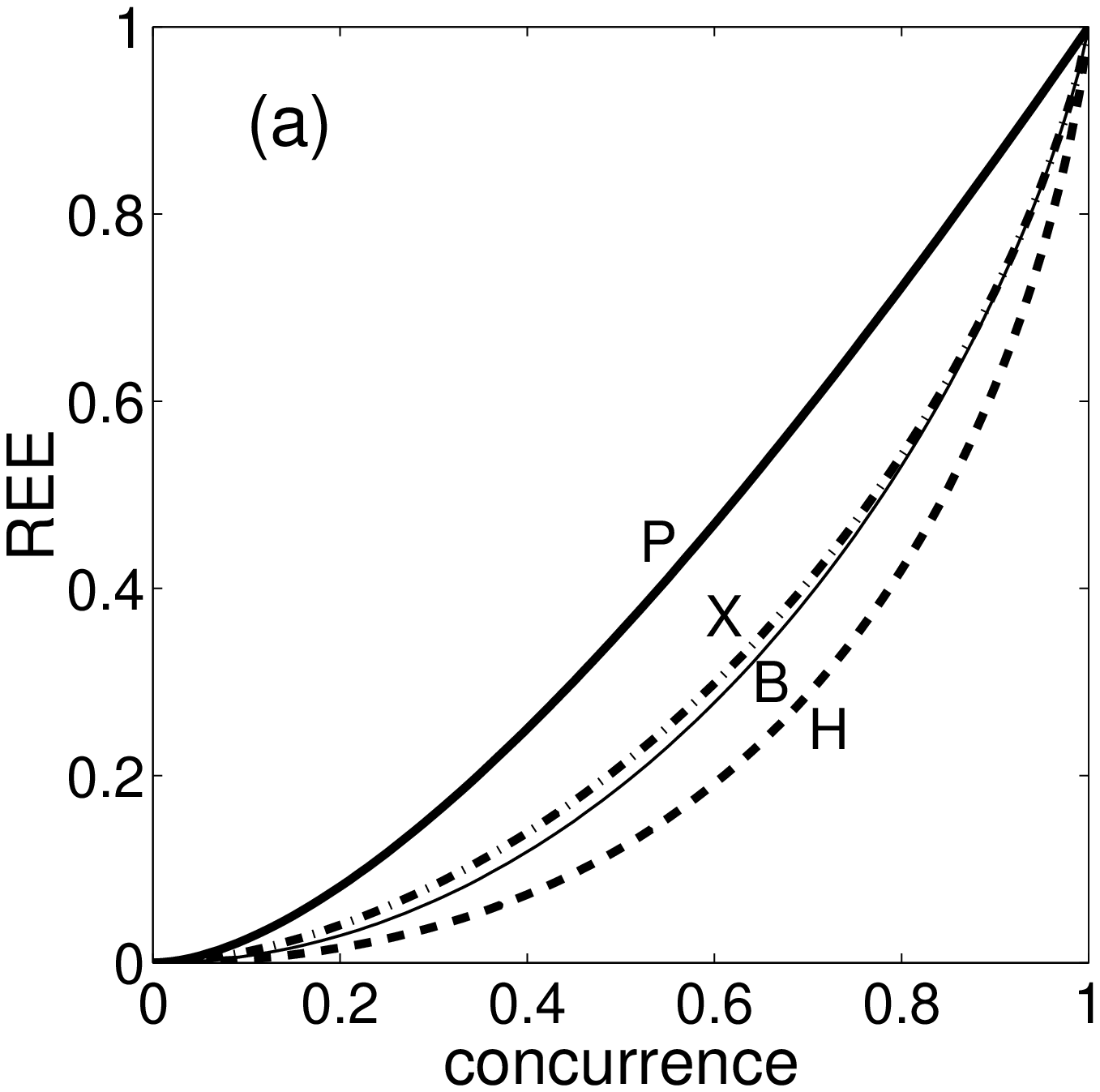}
 \epsfxsize=4cm\epsfbox{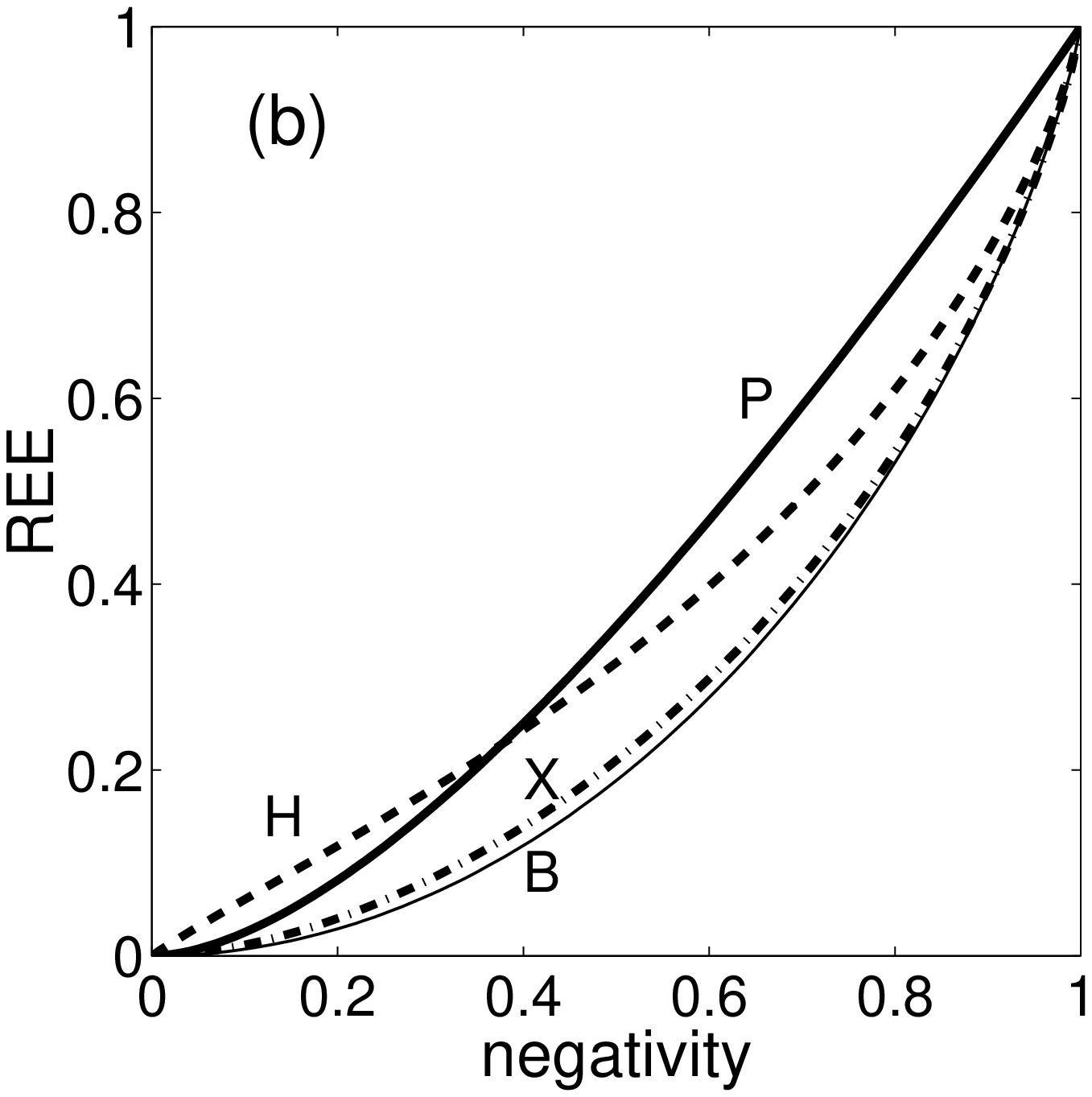}}
 \caption{REE versus (a) concurrence and (b)
negativity for the boundary states in figure 1(c).}
\end{figure}
\begin{figure}
\centerline{
 \epsfxsize=6cm\epsfbox{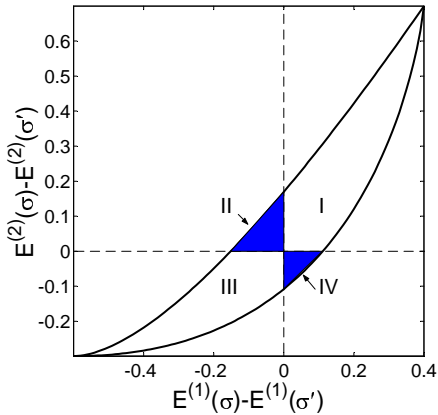}}
\caption{How to find states either satisfying or violating
condition (1): All states $\sigma$ for a given state $\sigma'$ for
which the chosen measures $E^{(1)}$ and $E^{(2)}$ impose the same
(opposite) order correspond to points in regions I and III (II and
IV). }
\end{figure}

The bounded regions containing all the generated states, as shown
in figure 1 and for clarity redrawn in figure 2, reveal the
ordering problem as a result of `the lack of precision with which
one entanglement measure characterizes the other' \cite{wei1}. By
simply generalizing the interpretation given by us in \cite{mg1} to
include any two ($E^{(1)}$ and $E^{(2)}$) of the studied
entanglement measures, one can conclude that for any partially
entangled state $\sigma'$ there are infinitely many partially
entangled states $\sigma$ for which the Eisert-Plenio condition,
given by (\ref{N01}), is violated. To demonstrate this result
explicitly for a given state $\sigma'$, it is useful to plot
$[E^{(2)}(\sigma)-E^{(2)}(\sigma')]$ versus
$[E^{(1)}(\sigma)-E^{(1)}(\sigma')]$ as shown in figure 3. Then the
state $\sigma$ corresponding to any point in the regions II and IV
is inconsistently ordered with $\sigma'$ with respect to the
measures $E^{(1)}$ and $E^{(2)}$. On the contrary, the states
$\sigma$, corresponding to any point in the regions I and III, and
$\sigma'$ are consistently ordered by $E^{(1)}$ and $E^{(2)}$.

Probability $P_{\rm ent}$ that a randomly generated two-qubit mixed
state is entangled can be estimated as $P_{\rm ent} \approx 0.368
\pm 0.002$ \cite{zyczkowski98} or $P_{\rm ent} \approx 0.365 \pm
0.001$ \cite{eisert}. However, probability $P_{\rm viol}$ that a
randomly generated pair of two-qubit states violates condition
(\ref{N01}) for concurrence and negativity is much less than
$P_{\rm ent}$ and estimated as $P_{\rm viol} \approx 0.047 \pm
0.001$ \cite{eisert}. Since the numerical analysis of Eisert and
Plenio \cite{eisert} and by the power of the Virmani-Plenio theorem
\cite{virmani} we know about the existence of states violating
condition (\ref{N01}). But it is not a trivial task to find
analytical examples of such states, especially in the case of the
orderings imposed by the REE in comparison to other entanglement
measures. We believe that it is not only a mathematical problem of
classification of states with respect to various entanglement
measures but it can shed more light on subtle physical aspects of
the entanglement measures including their operational
interpretation. By a comparison given in the next sections, we will
find states exhibiting very surprising properties. In particular,
we will show that states $\sigma'$ and $\sigma''$ can have the same
negativity, $N(\sigma')=N(\sigma'')$, the same concurrence,
$C(\sigma')=C(\sigma'')$, but still different REEs, $E(\sigma')\neq
E(\sigma'')$. A deeper analysis of such states can be useful in
studies of properties of a given entanglement measure (in this
example, the REE) under operations preserving other entanglement
measures (here, the entanglement of formation and the
PPT-entanglement cost). Thus, we believe that it is meaningful to
study analytically violation of condition (\ref{N01}) as will be
presented in greater detail in the next sections.

\section{Boundary states}

The extreme violation of (\ref{N01}) occurs if one of the states
corresponds to a point at the upper bound and the other at the
lower bound. Thus, for a comparison of different orderings, it is
essential to describe the states at the boundaries.

The upper bounds in figure 1 marked by $P$ correspond to two-qubit
pure states
\begin{equation}
|\psi_P \rangle =a|00\rangle +b|01\rangle +c|10\rangle
+d|11\rangle, \label{N07}
\end{equation}
where $a,b,c,d$ are the normalized complex amplitudes. The
concurrence and negativity are equal to each other and given by
\begin{equation}
C(|\psi _{P}\rangle)=N(|\psi _{P}\rangle) =2|ad-bc|. \label{N08}
\end{equation}
As shown by Verstraete {\em et al} \cite{verstraete}, the
negativity of any state $\sigma$ can never exceed its concurrence
[see figure 1(c)], and this bound is reached for the set of states
for which the eigenvector of the partial transpose of $\sigma$,
corresponding to the negative eigenvalue, is a Bell state.
Evidently, pure states belong to the Verstraete {\em et al} set of
states. For a pure state the REE is equal to the entanglement of
formation, thus is simply given by Wootters' relation (\ref{N04})
since $E(|{\psi}_{P}\rangle)=E_{\rm form}(|{\psi}_{P}\rangle)$. In
general, it holds $E_{\rm form}(\sigma)\ge E(\sigma)$
\cite{vedral98}, and the REE for pure states gives the upper bound
of the REE versus concurrence \cite{verstraete}. We have also
conjectured in \cite{mg2}, on the basis of numerical simulations
similar to those presented in figure 1(b), that the upper bound of
the REE versus negativity $N$ is reached by pure states for $N\ge
N_0\equiv 0.3770\cdots$.

Surprisingly, the REE versus $N$ for pure states, can be exceeded
by other states if $N<N_0$ as was shown in \cite{mg2} by the
so-called Horodecki states, which are mixtures of the maximally
entangled state, say the singlet state $|\psi_{-}\rangle
=(|01\rangle -|10\rangle )/\sqrt{2}$, and a separable state
orthogonal to it, say $|00\rangle$, i.e. \cite{horodecki-book}:
\begin{equation}
\sigma _{H}=C|\psi_{-}\rangle \langle \psi _{-}|+(1-C)|00\rangle
\langle 00| \label{N09}
\end{equation}
for which the concurrence and negativity are given, respectively,
by
\begin{subequations}
\begin{eqnarray}
C(\sigma _{H})&=&C, \label{N10a} \\
N(\sigma _{H})&=&\sqrt{ (1-C)^{2}+C^{2}}-(1-C). \label{N10b}
\end{eqnarray}
\end{subequations}
Verstraete {\em et al} \cite{verstraete} proved that a function of
the form (\ref{N10b}) determines the lower bound of the negativity
versus concurrence for any state $\sigma$ [see curve H figure
1(c)]. On the other hand, the REE versus concurrence for the
Horodecki states is given by \cite{vedral98}
\begin{equation}
E(\sigma _{H})=(C-2)\lg (1-C/2)+(1-C)\lg (1-C).\;\; \label{N11}
\end{equation}
By replacing $C$ by $\sqrt{2N(1+N)}-N$ in (\ref{N11}), one gets an
explicit dependence of $E(\sigma _{H})$ on the negativity $N(\sigma
_{H})$ \cite{mg2}. It was conjectured that the REE for the
Horodecki states describes the lower bound of the REE versus
concurrence \cite{vedral98}, as shown by curve H in figures 1(a)
and 2(a), and also conjectured \cite{mg2} that it gives the upper
bound of the REE versus negativity if $N\le N_0$ as seen in figures
1(b) and 2(b) \cite{mg2}. The ordering violation for any two of the
three entanglement measures can be shown for a pair of the
Horodecki and pure states, say $\sigma'$ and $\sigma$, if one of
the states is partially entangled ($0<E^{(1)}(\sigma')<1$) and
$\sigma$ is properly chosen according to the rule shown in figure 3
with an exception for the following case: If one of the states in
the pair of the Horodecki and pure states has the negativity equal
to $N_0$ then the ordering imposed by the REE and negativity for
these states is always consistent as required by condition
(\ref{N01}).

The lower bound in figure 1(b) and the upper bound figure 1(c)
correspond to the Bell diagonal state (labeled by $B$), given by
\begin{eqnarray}
\sigma _{B}=\sum_{i=1}^4 \lambda_i |\beta_{i} \rangle\langle
\beta_{i} | \label{N12}
\end{eqnarray}
with the largest eigenvalue $\max_j\lambda_j \equiv (1+C)/2\ge
1/2$, where $\sum_j\lambda_j=1$ and $|\beta_{i} \rangle$ are the
Bell states. The negativity and concurrence are the same and given
by
\begin{eqnarray}
C(\sigma _{B})= N(\sigma _{B})= C, \label{N13}
\end{eqnarray}
thus $\sigma _{B}$, similarly to pure states, belongs to the
Verstraete {\em et al} set of states maximizing the negativity for
a given concurrence. For the Bell diagonal states, the REE versus
the concurrence (and the negativity) reads as \cite{vedral97a}
\begin{eqnarray} \label{N14}
E(\sigma _{B})&=& 1-h((1+C)/2) \\
&=&\textstyle{\frac{1}{2}}\left[ (1+C)\lg (1+C)+(1-C)\lg
(1-C)\;\right]. \notag
\end{eqnarray}
If $\max_j\lambda_j\le 1/2$ then the state is separable, thus
$C(\sigma _{B})= N(\sigma _{B})=E(\sigma _{B})=0$. As an example of
(\ref{N12}), one can analyze the Werner state \cite{werner}
\begin{equation}
\sigma _{W}=\frac{1+2C}{3}|\psi _{-}\rangle \langle \psi
_{-}|+\frac{1-C}{6} I\otimes I,  \label{N15}
\end{equation}
where $0\le C\le 1$; $I$ is the identity operator of a single
qubit. Our choice of parametrization of (\ref{N01}) leads to
straightforward expressions for the negativity and concurrence
given by (\ref{N13}). The results of our simulation of $10^5$
random states presented in figure 1(b) confirm our conjecture in
\cite{mg2} that the lower bound of the REE versus negativity is
determined by the Bell diagonal states. Nevertheless, to our
knowledge, this conjecture and the other proposed by Verstraete
{\em et al} \cite{verstraete} on the lower bound of the REE versus
concurrence have not been proved yet \cite{horodecki04}. By
contrast, it is easy to prove, by applying local random rotations
to both qubits \cite{bennett96}, that the lower bound of the REE
versus fidelity is reached by the Bell diagonal states
\cite{vedral98}. It is worth noting that the REE versus concurrence
for $\sigma _{B}$ is not extreme as shown by curve $B$ in figure
2(a).

\begin{figure}
\centerline{
 \epsfxsize=5cm\epsfbox{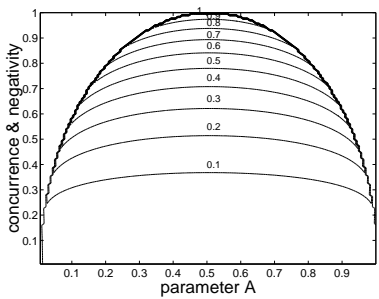}}
\caption{A contour plot of REE $E(\sigma_Y)$ as a function of
$C(\sigma_Y)=N(\sigma_Y)=C$ and parameter $A$ according to
(\ref{N21}).}
\end{figure}

Let us analyze another state corresponding to the upper bound for
$N$ versus $C$, but neither reaching the bounds for $E$ versus $C$
nor $E$ versus $N$. The state is defined as a MES, say the singlet
state, mixed with $|01\rangle$ as follows:
\begin{equation}
\sigma _{X}=C|\psi_{-}\rangle \langle \psi _{-}|+(1-C)|01\rangle
\langle 01| \label{N16}
\end{equation}
for which one gets
\begin{equation}
C(\sigma _{X})=N(\sigma _{X})=C. \label{N17}
\end{equation}
The eigenvalues of the partially transposed $\sigma_X$ are $\{
1-C/2,-C/2,C/2,C/2\}$ and they correspond to the eigenvectors given
by $\{|01\rangle,|\phi_{+}\rangle,|\phi_{-}\rangle,|10\rangle\}$,
where $|\phi_{\pm}\rangle=(|00\rangle\pm|11\rangle)/\sqrt{2}$.
Thus, the Verstraete condition for states with equal concurrence
and negativity is fulfilled for the state $\sigma _{X}$, as the
negative eigenvalue $-C/2$ corresponds to the Bell state. The
separable state $\bar{\rho}_X$ closest to $\sigma_X$ was found by
Vedral and Plenio \cite{vedral98} as
$\bar{\rho}_X=(1-C/2)|01\rangle\langle 01| +C/2|10\rangle\langle
10|$, which enables calculation of the following REE:
\begin{equation}
E(\sigma_X) = h(C/2)-h(r/2), \label{N18}
\end{equation}
where $r=1+\sqrt{(1-C)^2+C^2}$. Although (\ref{N17}) describes the
upper bound for $N$ versus $C$, (\ref{N18}) differs from the
extreme expressions for $E$ versus $C$ and $E$ versus $N$ given for
the pure, Horodecki and Bell diagonal states. Figures 2(a) and 2(b)
show clearly the differences.

We will also analyze the states dependent on two parameters defined
as
\begin{eqnarray}
\sigma_Y &=&A|01\rangle \langle 01|+(1-A)|10\rangle \langle 10|
\notag \\
&&+\frac{C}{2}(|01\rangle \langle 10|+|10\rangle \langle 01|)
\label{N19}
\end{eqnarray}
assuming that $C\leq 2\sqrt{A(1-A)}$ to ensure $\sigma_Y$ to be
positive semidefinite. States of the form, given by (\ref{N19}),
can be obtained by mixing a pure state $|\psi_P\rangle$ with the
separable state $\bar{\rho}_P$ closest to $|\psi_P\rangle$
\cite{mg2}. This mixing leaves the closest separable state
unchanged as implied by the Vedral-Plenio theorem \cite{vedral98}.
The eigenvalues of the partial transpose of $\sigma_Y$ are $\{1-A,
A,-C/2,C/2\}$, which correspond to the following eigenvectors
$\{|10\rangle,|01\rangle, |\phi_{-}\rangle,|\phi_{+}\rangle\}$,
respectively. Thus, the negative eigenvalue $-C/2$ corresponds to
the Bell state $|\phi_{-}\rangle$, which implies that $\sigma_Y$
belongs to the Verstraete {\em et al} set of states with equal
negativity and concurrence,
\begin{eqnarray}
C(\sigma_Y)=N(\sigma_Y)=C. \label{N20}
\end{eqnarray}
The REE for state (\ref{N19}) reads as
\begin{equation}
E(\sigma_Y)=h(A)-h\left(
\textstyle{\frac{1}{2}}[1+\sqrt{(1-2A)^2+C^{2}} ]\right)
\label{N21}
\end{equation}
which was obtained with the help of the closest separable state
$\bar{\rho}_P =A|01\rangle \langle 01|+(1-A)|10\rangle \langle 10|$
given in \cite{vedral98}. The contour plot of $E(\sigma_Y)$ is
shown in figure 4. The states (\ref{N19}), independent of parameter
$A$, are the upper bound states for $N$ versus $C$. By changing
$A$, the states (\ref{N19}) transform from the pure states into
Bell diagonal states, thus they can become the upper bound states
both for $E$ versus $C$ and $E$ versus $N\ge N_0$, as well as the
lower bound states for $E$ versus $N$. In general, a state
corresponding to any point between curves $P$ and $B$ in figures
2(a) and 2(b) can be given by (\ref{N19}).

\section{Analytical comparison of state orderings}

By analyzing pairs of states discussed in the previous section and
by applying the rule shown in figure 3 we can easily find
analytical explicit examples of states violating condition
(\ref{N01}) by any two measures out of the triple, when the third
measure is not analyzed. However, the number of classes of state
pairs increases to 14, as shown in table 1, on including all
possible different predictions of the state orderings imposed by
all the three measures simultaneously. The number of classes is
given mathematically by permutation with replacement (where the
order counts and repetitions are allowed) and equal to $3^3$. But
we should not count twice the classes defined by opposite
inequalities [e.g., class 2 can be equivalently given by $C(\sigma
')>C(\sigma '')$, $N(\sigma ')<N(\sigma '')$, $E(\sigma ')>E(\sigma
'')$] since the definition of states $\sigma '$ and $\sigma''$ can
be interchanged. Thus, the number of classes decreases to
$(3^3-1)/2+1=14$. One can identify all these classes by analyzing
pairs of points in the crescent-like solid region in $CNE$ space
shown in figure 5 with the familiar projections into the planes
$CE$ [see also figure 1(a)], $NE$ [figure 1(b)], and $CN$ [figure
1(c)]. Unfortunately, a graphical illustration of various cross
sections of the solid crescent in figure 5 would not be clear
enough. Thus, in figure 6, we give a symbolic representation of the
14 classes of table 1 by depicting only small cubes around  point
$[C(\sigma'),N(\sigma'),E(\sigma')]$ for a given state $\sigma'$.
In a sense, the cubes are cut inside the solid crescent shown in
figure 5.

In the following, we will give explicit examples of the pairs of
states satisfying the inequalities listed in table 1. For compact
notation we denote
\begin{equation*}
\Delta\equiv [C(\sigma'')-C(\sigma'),
N(\sigma'')-N(\sigma'),E(\sigma'')-E(\sigma')].
\end{equation*}
States consistently ordered by all the three measures as required
by the Eisert-Plenio condition (\ref{N01}) belong to {\em class 1}.
The vast majority of the randomly generated pairs of two-qubit
states belong to this class. The simplest analytical example is a
pair of pure states $|\psi_{i}\rangle=a_i|00\rangle +b_i|01\rangle
+c_i|10\rangle +d_i|11\rangle$ ($i$=1,2), for which
$|a_1d_1-b_1c_1|\neq |a_2d_2-b_2c_2|$. Similarly, by comparing
other pairs of states, to mention $(\sigma_H(C'),\sigma_H(C''))$,
$(\sigma_B(C'),\sigma_B(C''))$ or $(\sigma_X(C'),\sigma_X(C''))$
for $C'\neq C''$, one arrives at the same conclusion. A pair of
states from {\em class 2} can be given, e.g., by the Bell diagonal
and Horodecki states for slightly different concurrences (or
negativities). E.g., if $\sigma_B(C=0.5)$ and $\sigma_H(C=0.6)$
then $\Delta=[0.1,-0.179,0.003]$, or for the same $\sigma_B$ but
$\sigma_H$ having its negativity equal to $0.4$ then
$\Delta=[0.158,-0.1,0.055]$ as required. As an example of the state
pair from {\em class 3}, we choose the Horodecki and pure states
such that their negativities are close to $N_0$. E.g., let
$\sigma_H$ have the negativity $N_0-0.1$ and $|\psi_P\rangle$ have
its coefficients satisfying $2|ad-bc|=N_0$ then
$\Delta=[-0.187,0.1,0.064]$. By choosing pure state with
concurrence $C'=0.625\cdots$ and the Horodecki state for
$C''=0.846\cdots\equiv C_0$, we observe that their REEs are the
same. Then, an example of the state pair from {\em class 4} can be
given by the above pure state and the Horodecki state with its
concurrence slightly less than $C_0$, say $C(\sigma_H)=C_0-0.02$,
which implies that $\Delta=[0.200,0.044,-0.037]$ as required. The
classes 1--4 are defined solely by sharp inequalities, and thus
they are crucial in our comparison of different state orderings.
\begin{table} 
\caption{All possible different predictions of the state orderings
imposed by the REE, concurrence and negativity. As explained in the
text, the remaining 13 classes of state pairs can be obtained by
the listed classes just by interchanging definitions of $\sigma'$
and $\sigma''$. Asterisk denotes the classes for which we were not
able to find examples.}
\begin{center}
\begin{tabular}{l l l l}
\hline
Class \hspace{1mm} & Concurrences \hspace{4mm} & Negativities \hspace*{7mm} & REEs \\
\hline
1 & $C(\sigma')<C(\sigma'')$, & $N(\sigma')<N(\sigma'')$, & $E(\sigma')<E(\sigma'')$\\
2 &$ C(\sigma')<C(\sigma'')$, & $N(\sigma')>N(\sigma'')$, & $E(\sigma')<E(\sigma'')$\\
3 & $C(\sigma')>C(\sigma'')$, & $N(\sigma')<N(\sigma'')$, & $E(\sigma')<E(\sigma'')$\\
4 & $C(\sigma')<C(\sigma'')$, & $N(\sigma')<N(\sigma'')$, & $E(\sigma')>E(\sigma'')$\\
\vspace{1mm}
5 & $C(\sigma')=C(\sigma'')$, & $N(\sigma')=N(\sigma'')$, & $E(\sigma')=E(\sigma'')$\\
6 & $C(\sigma')<C(\sigma'')$, & $N(\sigma')=N(\sigma'')$, & $E(\sigma')<E(\sigma'')$\\
7 & $C(\sigma')=C(\sigma'')$, & $N(\sigma')<N(\sigma'')$, & $E(\sigma')<E(\sigma'')$\\
8 & $C(\sigma')<C(\sigma'')$, & $N(\sigma')<N(\sigma'')$, & $E(\sigma')=E(\sigma'')$\\
9 & $C(\sigma')=C(\sigma'')$, & $N(\sigma')=N(\sigma'')$, & $E(\sigma')<E(\sigma'')$\\
10 & $C(\sigma')<C(\sigma'')$, & $N(\sigma')=N(\sigma'')$, & $E(\sigma')=E(\sigma'')$\\
11$^*$ & $C(\sigma')=C(\sigma'')$, & $N(\sigma')<N(\sigma'')$, & $E(\sigma')=E(\sigma'')$\\
12$^*$ & $C(\sigma')>C(\sigma'')$, & $N(\sigma')=N(\sigma'')$, & $E(\sigma')<E(\sigma'')$\\
13$^*$ & $C(\sigma')=C(\sigma'')$, & $N(\sigma')>N(\sigma'')$, & $E(\sigma')<E(\sigma'')$\\
14 & $C(\sigma')<C(\sigma'')$, & $N(\sigma')>N(\sigma'')$, & $E(\sigma')=E(\sigma'')$\\
\hline
\end{tabular}
\end{center}
\end{table}
\begin{figure} 
\epsfxsize=6cm\centerline{\epsfbox{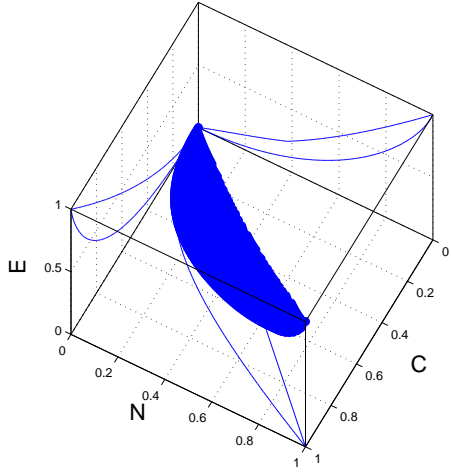}} \caption{States
$\sigma$ characterized by $[C(\sigma),N(\sigma),E(\sigma)]$ lie in
the solid crescent-like region with its projections into the planes
shown in figure 1. All classes of state pairs from table 1 can be
found by analyzing pairs of points at various cross sections of the
region.}
\end{figure}
\begin{figure} 
\epsfxsize=9cm\centerline{\epsfbox{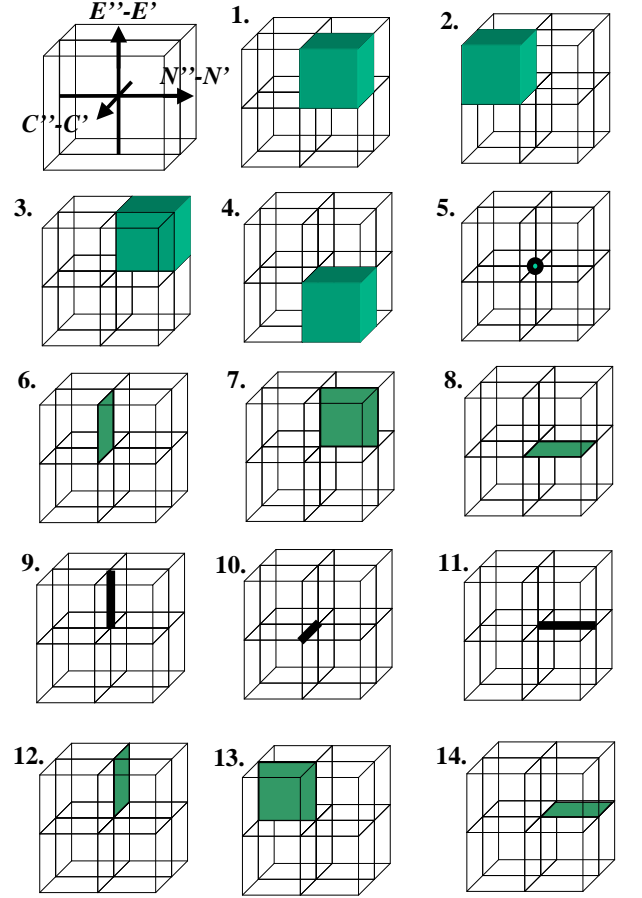}} \caption{A
schematic representation of the 14 classes of state pairs listed in
table 1, where $f'=f(\sigma')$ and $f''=f(\sigma'')$ for $f=C,N,E$.
The central point corresponds to a state $\sigma'$ for which
$\Delta=[0,0,0]$. A pair of states $\sigma'$ and $\sigma''$, where
the latter is represented by any point inside the marked region of
the $i$th ($i=1,...,14$) sub-figure, satisfies the inequalities of
the $i$th class in table 1.}
\end{figure}

Now, we will present more subtle comparison to include the classes,
when some of the entanglement measures are equal to each other for
different states. {\em Class 5} is interesting enough to be
analyzed separately in the next section. An example of the state
pair from {\em class 6} can be given by the Bell diagonal and
Horodecki states with the same negativities, say equal to $1/2$,
which implies that $\Delta=[0.225,0,0.127]$. Also a member of {\em
class 7} can be given by the above states but for the same
concurrences, say $C=0.5$, which implies that
$-\Delta=[0,0.293,0.066]$. Simple examples of the state pairs from
{\em classes 6} and {\em 7} can also be found by considering the
following state:
\begin{eqnarray}
\sigma_Z(C,N) =\textstyle{\frac{1}{2}}[(1-\alpha )(|01\rangle
\langle
01|+|10\rangle \langle 10|)  \notag \\
+C(|01\rangle \langle 10|+|10\rangle \langle 01|)+2\alpha
|00\rangle \langle 00|]  \label{N26}
\end{eqnarray}
for $N>0$ and $C\in\langle N,\sqrt{2N(N+1)}-N\rangle$, where
$\alpha =(C^{2}-N^{2})/(2N)$. The range-limited $C$ ensures
semi-definiteness of $\sigma_Z$. State (\ref{N26}) can be generated
by mixing the Horodecki state $\sigma_H$ with the separable state
$\bar{\rho}_H$ closest to $\sigma_H$ given by Vedral and Plenio
\cite{vedral98} (for details see \cite{mg2}). We note that the
coefficients $C$ and $N$ in (\ref{N26}) are chosen so that
\begin{equation}
C(\sigma_Z)=C, \quad N(\sigma_Z)=N. \label{N27}
\end{equation}
Then, we can write the REE as follows:
\begin{eqnarray}
E(\sigma_Z) &=& h_3\left( (1+\alpha )\beta ,\textstyle{\frac{1}{2}}
(1+\alpha )(1-2\beta )+\beta C\right) \notag \\
&&- h_3\left( \alpha ,\textstyle{\frac{1}{2}}(1-\alpha +C)\right),
\label{N28}
\end{eqnarray}
where $\beta =\alpha (1+\alpha )/[(1+\alpha )^{2}-C^{2}]$ and
$h_3(x_{1},x_{2})=-\sum_{i=1}^3 x_{i}\lg x_{i}$ with
$x_3=1-x_1-x_2$. By changing $C$ and $N$ separately, we can obtain
$\sigma_Z$ with a desired REE. For example, by fixing the
negativity, we get the state pair corresponding to {\em class 6},
as shown by the contours of constant negativity in figure 7(a). On
the other hand, by fixing the concurrence, the resulting states
$\sigma_Z$ satisfy the conditions for {\em class 7}, as presented
by the contours of constant concurrence in figure 7(b).
\begin{figure}
\centerline{
 \epsfxsize=4cm\epsfbox{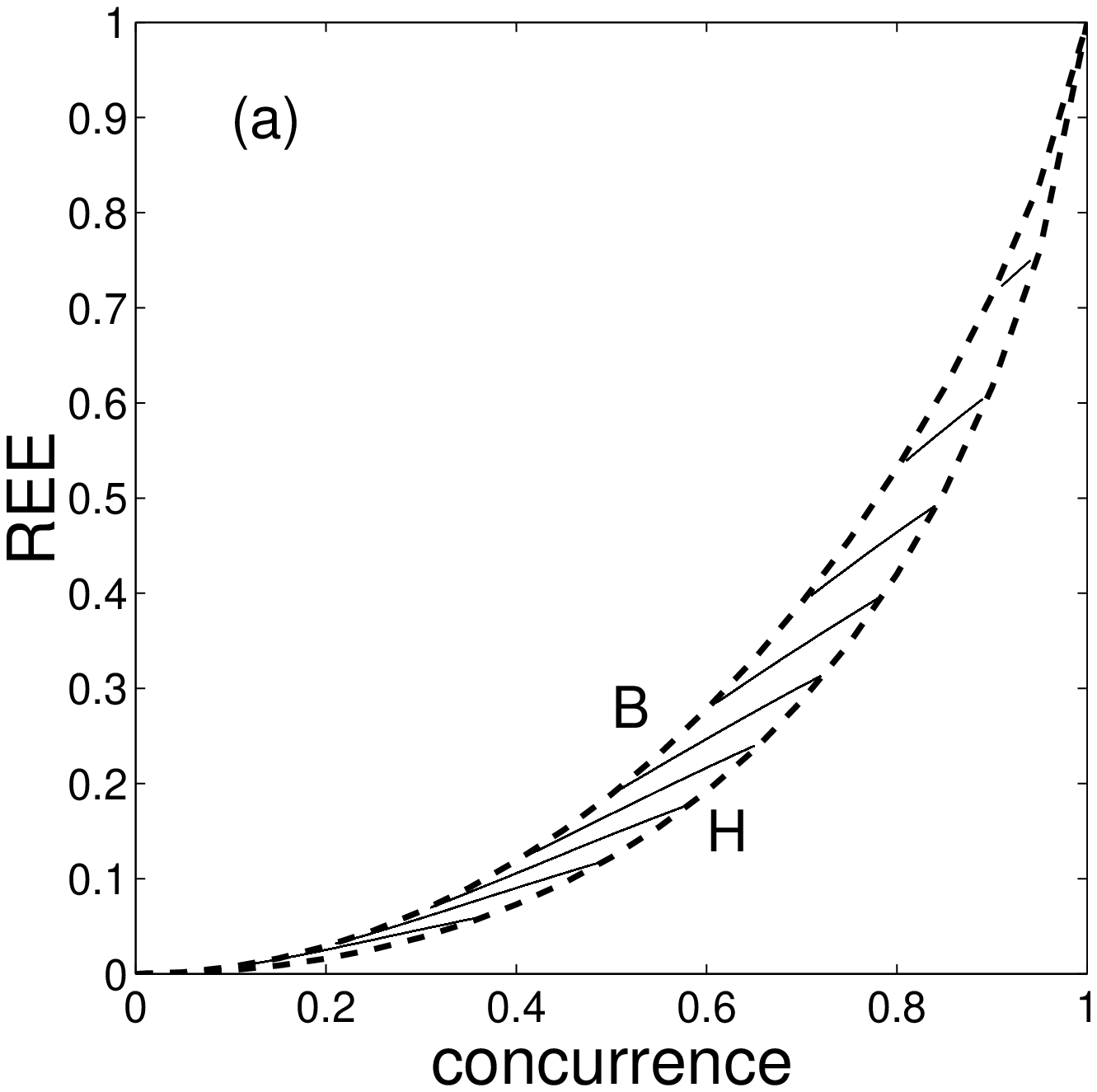}
 \epsfxsize=4cm\epsfbox{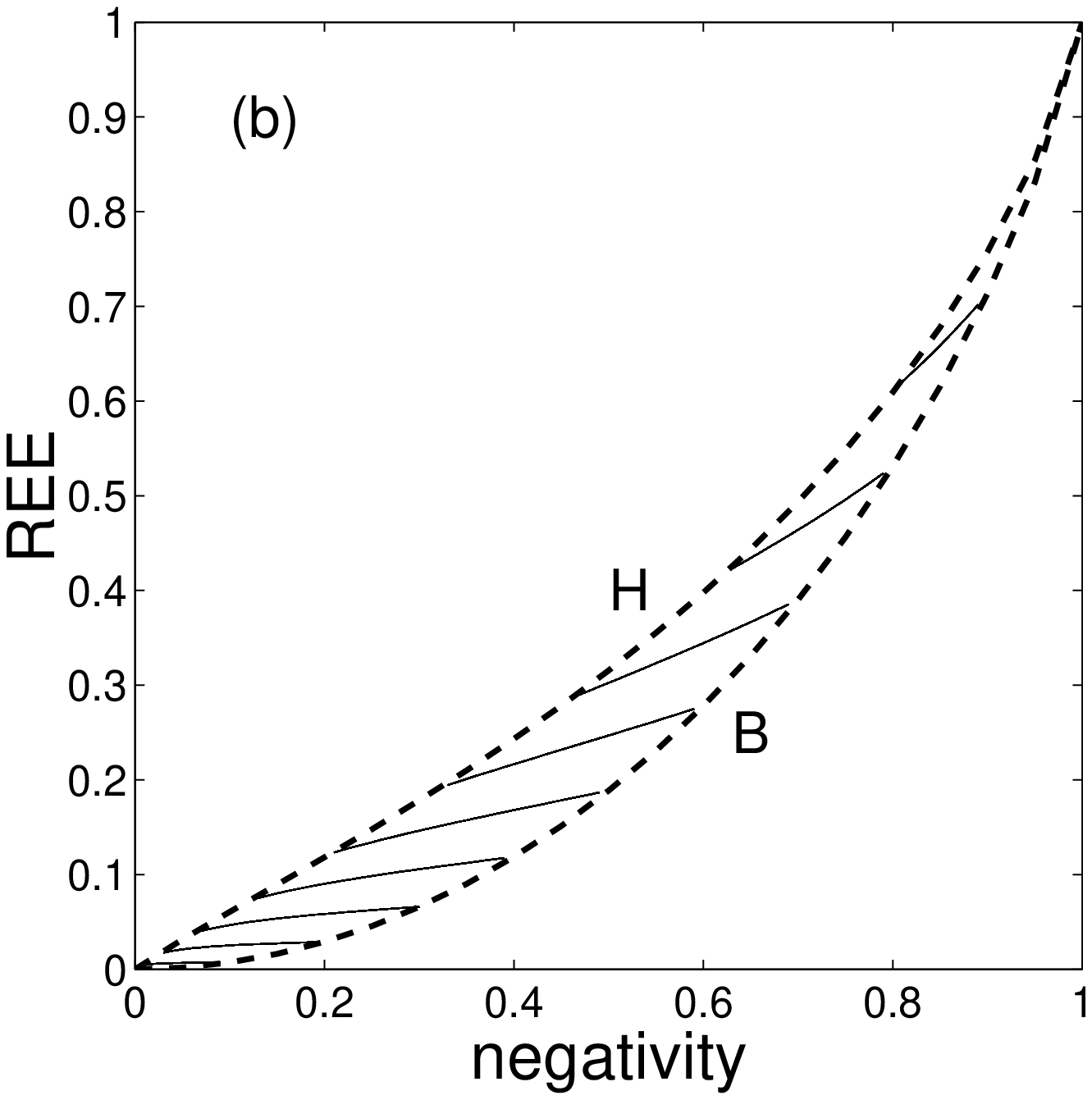}}
 \centerline{\epsfxsize=4cm\epsfbox{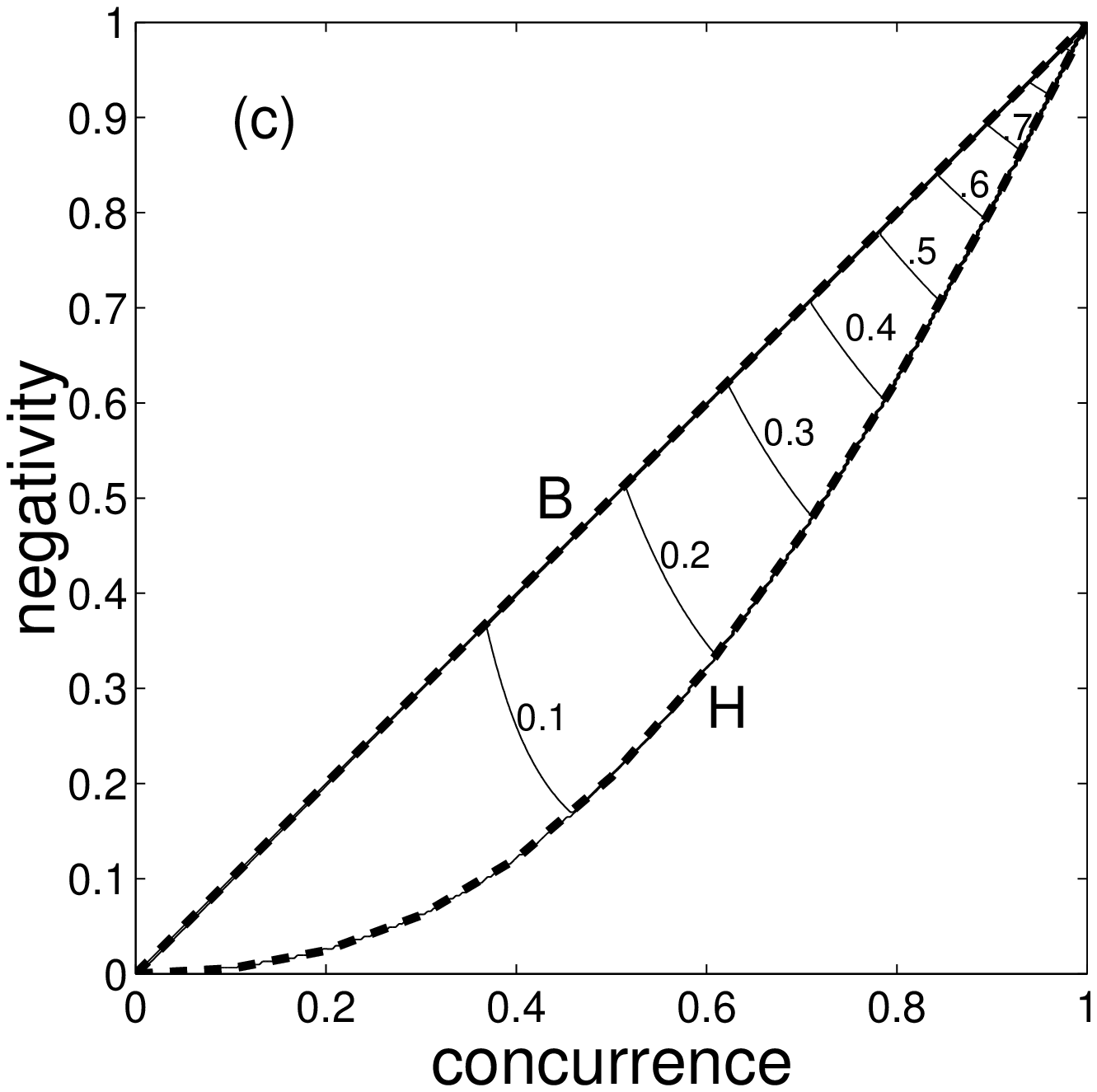}}
\caption{Contour plots of the entanglement measures for $\sigma_Z$:
(a) negativity $N(\sigma_Z)$ as a function of $C(\sigma_Z)$ and
$E(\sigma_Z)$, (b) concurrence $C(\sigma_Z)$ as a function of
$N(\sigma_Z)$ and $E(\sigma_Z)$, and (c) REE $E(\sigma_Z)$ as a
function of $C(\sigma_Z)$ and $N(\sigma_Z)$. The contours are
depicted at values of $0.1,0.2,...1$ from the left bottom corner to
right upper corner.}
\end{figure}

To {\em class 8} belongs a pair of, e.g., the pure state with
concurrence $0.625\cdots$ and the Horodecki state with
$C=0.846\cdots$, then it holds $E(|\psi_P\rangle)=E(\sigma_H)=0.5$,
and $\Delta=[0.220,0.080,0]$ as requested. To find an exemplary
member of {\em class 9}, one can compare a pure state and any other
state from the Verstraete {\em et al} set of states (including
$\sigma_B$, $\sigma_X$ or $\sigma_Y$) with the same concurrence,
which means also the same negativity. For example, for
$C(|\psi_P\rangle)=C(\sigma_B)=1/2$ one gets $\Delta=[0,0,0.189]$.
As regards {\em class 10}, we can compare the pure and Horodecki
states with the same negativity $N=N_0$, which implies that
$E(|\psi_P\rangle)=E(\sigma_H)$. Thus, we have
$\Delta=[0.265,0,0]$. Unfortunately, by comparing the states
discussed in this section, we have not found examples of the state
pairs from {\em classes 11--13}. But we can give a few exemplary
members of {\em class 14}. E.g., by comparing the Bell diagonal
state for $C'=0.779\cdots$ and the Horodecki state for
$C''=0.846\cdots$ we find that $E(\sigma_B)=E(\sigma_H)=0.5$, while
their negativities and concurrences violate condition (\ref{N01})
to the following degrees $\Delta=[0.066,-0.074,0]$. Also by
analyzing figure 7(c) for any two points at the same contour of
constant REE, we find exemplary state pairs from {\em class 14}.
Thus, we have presented simple analytical examples of the states
satisfying 11 out of 14 classes listed in table 1.

\section{States with the same $E$, $C$ and $N$}

Here, we will analyze examples of inequivalent states
$\sigma'\neq\sigma''$, which have the same degree of entanglement
according to $E$, $C$, and $N$, thus corresponding to {\em class 5}
in table 1. It is tempting to choose simply two different pure
states with their coefficients satisfying $|a_1d_1-b_1c_1|=
|a_2d_2-b_2c_2|$, which guarantees the fulfillment of the
equalities required for this class. However, such pure states can
be transformed into each other by local operations. To show this,
first we note that any pure state, given by (\ref{N07}), can be
transformed by local rotations into the superposition
$|\tilde{\psi}_P(p) \rangle =\sqrt{p} |01\rangle + \sqrt{1-p}
|10\rangle$ ($0\le p\le 1$), for which the concurrence and
negativity are equal to $2\sqrt{p(1-p)}$, as a special case of
(\ref{N08}). The same value of these entanglement measures occurs
also for $|\tilde{\psi}_P(1-p) \rangle$, but this state can be
transformed into $|\tilde{\psi}_P(p) \rangle$ by applying NOT gate
to each of the qubits. Thus, we have shown that pure states are not
good examples of the state pairs from {\em class 5}. Then, let us
choose, e.g., two different Bell diagonal states but with the same
largest eigenvalue greater than $1/2$. By virtue of (\ref{N13}) and
(\ref{N14}), we conclude that these states have the same degree of
entanglement according to the REE, concurrence and negativity.
However, as we will show in the following, they can violate the
Bell inequality to different degrees.

The maximum possible violation of the Bell inequality in the
Clauser-Horne-Shimony-Holt (CHSH) form \cite{clauser}
\begin{equation}
|\langle {\cal B}\rangle_{\sigma}|=|{\cal E}(\phi_1,\phi_2)+{\cal
E}(\phi_1',\phi_2)+{\cal E}(\phi_1,\phi_2')-{\cal
E}(\phi_1',\phi_2')|\leq 2 \label{N22}
\end{equation}
for a two-qubit state $\sigma$ is given by \cite{horodecki95}
\begin{equation}
\max_{\cal B}\langle {\cal B}\rangle_{\sigma}=2\sqrt{M(\sigma)}.
\label{N23}
\end{equation}
Here, ${\cal B}$ is the Bell operator, $\phi_i,\phi_i'$ are two
dichotomic variables of the $i$th qubit, and ${\cal
E}(\phi_1,\phi_2)$ is the expectation value of the joint
measurement of $\phi_1$ and $\phi_2$, and so on for the other
expectation values. The quantity $M(\sigma)$ is the sum of the two
largest eigenvalues of $T_{p}T_{p}^{\dagger}$, where $T_{p}$ is the
$3\times3$ matrix formed by the elements $t_{nm}={\rm
Tr}(\sigma\sigma^{(n)} \otimes\sigma^{(m)})$ given in terms of the
Pauli matrices $\sigma^{(j)}$. Inequality (\ref{N22}) is satisfied
if and only if $M(\sigma)\le 1$ \cite{horodecki95}. As shown in
\cite{miran2} for any pure state $|\psi_P\rangle$, the Bell
inequality violation parameter $M(\sigma)$ is closely related to
the concurrence and negativity as follows:
\begin{equation}
\sqrt{ \max \,\{0,\,M(|\psi_P\rangle)-1\,\}}
=C(|\psi_P\rangle)=N(|\psi_P\rangle). \label{N24}
\end{equation}
We find that $M(\sigma)$ for the Bell diagonal state reads as
\begin{equation}
M(\sigma_B)=2\max_{(i,j,k)}[(\lambda_i-\lambda_j)^2
+(\lambda_k-\lambda_4)^2], \label{N25}
\end{equation}
where subscripts $(i,j,k)$ change over cyclic permutations of
$(1,2,3)$. Concluding, the Bell-inequality violation depends on all
$\lambda_i$'s, while the entanglement measures $E$, $C$, and $N$
depend solely on the largest $\lambda_i$. Thus, as an example of
the state pair from {\em class 5}, we can choose two Bell diagonal
states $\sigma'_{B}$ and $\sigma''_{B}$ with only the largest
eigenvalue being the same and greater than $1/2$ for both states,
which implies that the states cannot be transformed into each other
by LOCC operations but still have the same degrees of entanglement:
$E(\sigma'_{B})=E(\sigma''_{B})$, $C(\sigma'_{B})=C(\sigma''_{B})$
and $N(\sigma'_{B})=N(\sigma''_{B})$.

\section{Conclusions}

We have analyzed the problem of inconsistency in ordering states
with the entanglement measures. The problem was raised by Eisert
and Plenio \cite{eisert} on the numerical example of the
concurrence and negativity and then studied by others
\cite{zyczkowski99,verstraete,zyczkowski02,wei1,wei2,miran1,miran2,mg1}.
The ordering problem is closely related to existence of the upper
and lower bounds of one entanglement measure versus the other
\cite{verstraete,wei1,mg1,mg2}. Here, we presented analytical
examples of the pairs of states consistently and inconsistently
ordered by the relative entropy of entanglement in comparison to
the concurrence and negativity. In particular, we have found
examples of the states for which any of the measures imposes order
opposite to that given by the other two measures, which corresponds
to {\em classes 1--4} in table 1. We have also identified pairs of
states with, in particular, (i) the same concurrences and
negativities but different REEs (as corresponding to {\em class
9}), (ii) the same REEs and negativities but different concurrences
({\em class 10}), (iii) the same REEs but different and oppositely
ordered concurrences and negativities ({\em class 14}), or (iv)
states having the same three entanglement measures ({\em class 5}),
but still violating the Bell-CHSH inequality to different degrees.

{\bf Acknowledgments}. We are grateful to M~Horo\-decki,
P~Horodecki, Z Hradil, G Kimura, W Leo\'nski, R Tana\'s, F
Verstraete, S Virmani, A W\'ojcik and K \.Zyczkowski for their
valuable comments. AG was supported in part by the Polish State
Committee for Scientific Research, Contract No.~0~T00A 003 23.

\end{document}